\documentclass[11pt]{article}

\usepackage{euscript}
\usepackage{amssymb}
\usepackage{amsfonts}
\usepackage{amsbsy}
\usepackage{amsmath}
\usepackage{epsfig}
\usepackage{amsthm}
\usepackage{amscd}
\usepackage{amstext}
\usepackage{verbatim}

\usepackage[pdftex]{hyperref}

\def\ben{\begin{equation}}
\def\een{\end{equation}}
\def\half{{\textstyle{1\over2}}}

\let\a=\alpha   \let\d=\delta 
  \let\q=\theta 
\let\l=\lambda     
 \let\t=\tau

\let\w=\omega

\let\pa=\partial
\def\be{\begin{equation}}
\def\ee{\end{equation}}
\def\ba{\begin{array}}
\def\ea{\end{array}}

\def\dalemb#1#2{{\vbox{\hrule height .#2pt
       \hbox{\vrule width.#2pt height#1pt \kern#1pt
               \vrule width.#2pt}
       \hrule height.#2pt}}}

\newcommand{\bea}{\begin{eqnarray}}
\newcommand{\eea}{\end{eqnarray}}

\def\Z{{{\Bbb Z}}}

\def\ocal{{\mathcal{O}}}

\numberwithin{equation}{section}

\begin{document}

\begin{flushright} MIT-CTP 4148 \\ \end{flushright}
\vspace{1cm}

\begin{center}

{ \LARGE {\bf Holography and the \\ Coleman-Mermin-Wagner theorem}}

\vspace{1.2cm}

Dionysios Anninos$^\flat$, Sean A. Hartnoll$^\flat$ and Nabil Iqbal$^\sharp$

\vspace{0.9cm}

{\it $^\flat$ Center for the Fundamental Laws of Nature, Harvard University,\\
Cambridge, MA 02138, USA \\}

\vspace{0.5cm}

{\it $^\sharp$ Center for Theoretical Physics, Massachusetts Insitute of Technology,\\
Cambridge, MA 02139, USA \\}

\vspace{0.5cm}

{\tt anninos@physics.harvard.edu, hartnoll@physics.harvard.edu, niqbal@mit.edu} \\

\vspace{1.3cm}

\end{center}

\begin{abstract}

In 2+1 dimensions at finite temperature, spontaneous symmetry breaking of global symmetries is precluded by large thermal fluctuations of the order parameter. The holographic correspondence implies that analogous effects must also occur in 3+1 dimensional theories with gauged symmetries in certain curved spacetimes with horizon. By performing a one loop computation in the background of a holographic superconductor, we show that bulk quantum fluctuations wash out the classical order parameter at sufficiently large distance scales. The low temperature phase is seen to exhibit algebraic long range order. Beyond the specific example we study, holography suggests that IR singular quantum fluctuations of the fields and geometry will play an interesting role for many 3+1 dimensional asymptotically AdS spacetimes with planar horizon.

\end{abstract}

\pagebreak
\setcounter{page}{1}

\section{Introduction}

The Coleman-Mermin-Wagner-Hohenberg (CMWH) theorem \cite{sym} states that continuous global symmetries cannot
be broken spontaneously in 1+1 dimensions at zero temperature or in 2+1 dimensions at finite temperature.
The rigorous proofs in these papers implement the long appreciated (e.g. \cite{bloch}) physical fact that divergent
infrared thermal or quantum fluctuations of the putative order parameter wash out classical expectation
values in low dimensions.

The holographic correspondence \cite{Maldacena:1997re} relates quantum field theories in d+1 dimensions to
gravitational theories in one higher spatial dimension. The correspondence often admits a 
`large $N$' limit in which the dual gravitational theory may be treated semiclassically. In this limit
the strongly interacting quantum field theory is dually described by fields propagating classically on a fixed
curved spacetime background in one higher dimension. Each field in the `bulk' corresponds to an operator
in the `boundary' quantum field theory. The bulk description
of spontaneous symmetry breaking is a field charged under the symmetry which has a nonzero purely normalisable\footnote{Near the boundary, the bulk field is characterised by two modes:
a `normalisable' mode that can be excited with finite energy and `non-normalisable' mode that is fixed by
the boundary conditions. Given a bulk field configuration, the normalisable mode determines the
expectation value of the dual operator while the presence of a non-normalisable mode indicates that
a source for this operator has been added to the action of the theory.}
behaviour near the boundary \cite{Klebanov:1999tb}.

The combination of holography and the CMWH theorem implies that there exist spacetime geometries in one
higher dimension than those appearing in the statement of the theorem above in which spontaneous symmetry
breaking is prevented by strong infrared fluctuations. A further difference with the usual CMWH theorem is that the bulk symmetry in question is gauged rather than global.
The main objective of this paper is to explicitly demonstrate
this phenomenon via a bulk one loop computation. This will allow us to show that
the phase of the order parameter is washed out everywhere in the bulk rather than only near the asymptotic boundary. It will
also enable us to exhibit the expected algebraic long range order in the low temperature phase. In general it is difficult
to compute loop effects in curved spacetimes, where field propagators cannot be found exactly.
Our work is in the spirit of recent computations \cite{Denef:2009yy, CaronHuot:2009iq, Hartnoll:2009kk, Faulkner:2010da, Hartman:2010fk}
in `applied holography' \cite{Hartnoll:2009sz, McGreevy:2009xe} showing how quantum effects due to gapless modes running in loops can be accurately captured.

A one loop bulk computation is required because in the classical bulk limit the number of degrees of freedom
`per site' in the dual field theory becomes very large. In this `large $N$' limit, the dangerous fluctuations about the classical `mean field' value
are parametrically suppressed. This effect was demonstrated explicitly in a field theoretic computation for the $SU(N)$ Thirring model in \cite{Witten:1978qu}. We will recall more precisely below the sense in which the classical limit correctly captures the large $N$ physics, and therefore the sense in which previous classical computations using 2+1 dimensional holographic superconductors are correct at large but finite $N$. While interesting physics resembling the CMWH theorem has been reported within a classical gravitational framework \cite{Marolf:1999uq, Peet:1999bc}, the fact that only saddle point computations were required in the bulk indicates that those effects were not fluctuation driven in the sense that we mean it. Similarly, adding higher derivative `$1/N$' terms to the bulk action \cite{Gregory:2009fj} does not move one off a saddle point: local bulk counterterms can capture high energy (UV) quantum corrections while the CMWH theorem is a nonanalytic low energy (IR) effect.

Our computation will closely mirror the standard discussions of one loop corrections to classical expectation values in flat space quantum field theory. A three point interaction in a classical background with broken symmetry will lead to a tadpole diagram describing the one loop correction. Sending a hydrodynamic sound mode around the loop produces an IR divergent contribution that washes out the phase of the classical order parameter and restores the symmetry. This process is illustrated in figure \ref{fig:oneloop} below. Our work is complicated by the coupling between several fields and by the need to fix a gauge, but in essence is straightforward.

\begin{figure}[h]
\begin{center}
\includegraphics[height=180pt]{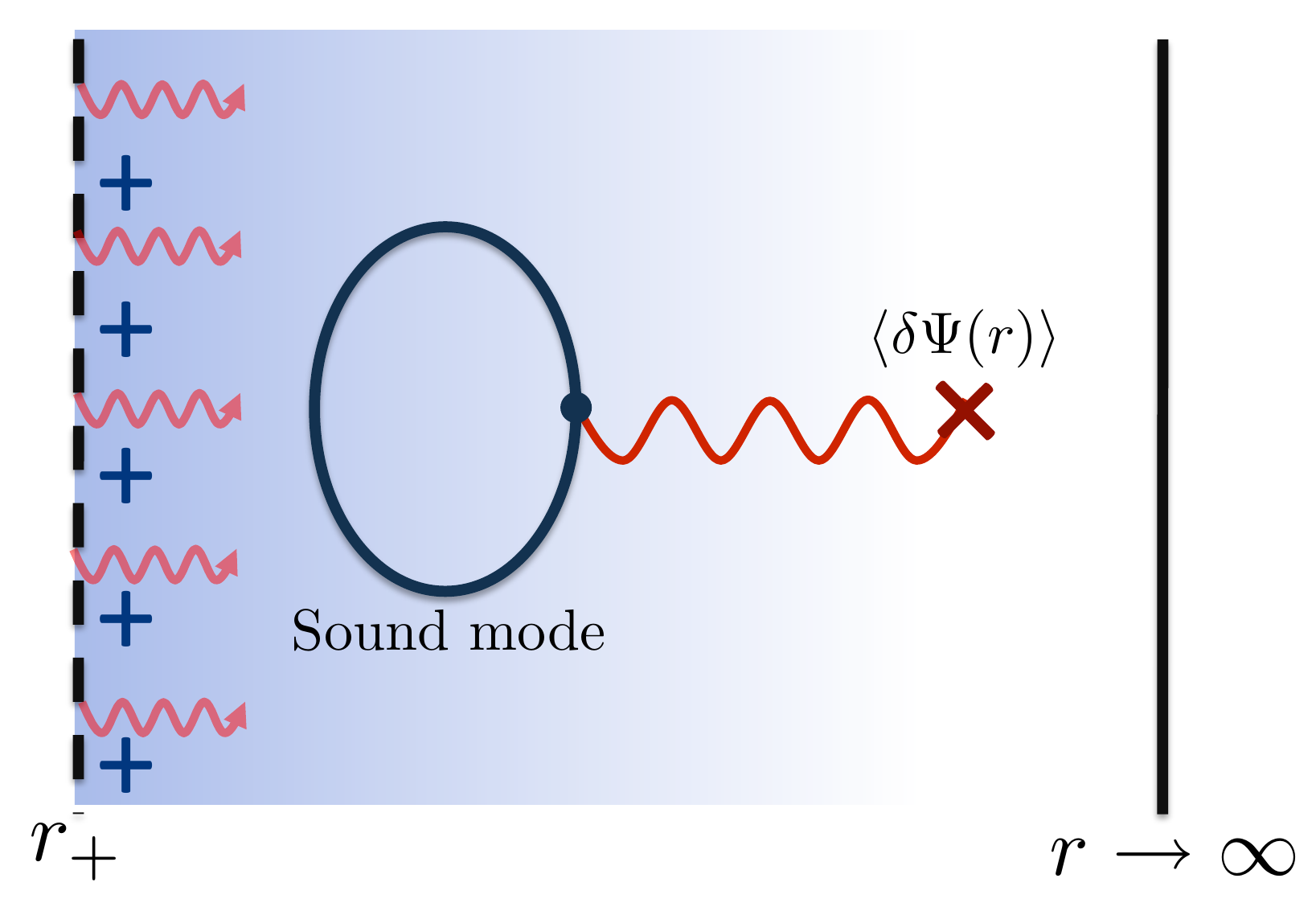}
\end{center}
\caption{The classical background is a planar AdS$_4$ black hole, carrying charge and surrounded by symmetry breaking scalar `hair'. We exhibit the divergent tadpole contribution of a hydrodynamic sound mode that randomizes the phase of the order parameter.}\label{fig:oneloop}
\end{figure}

We will illustrate the physics of the bulk CMWH theorem by working through a particular simple example: holographic superconductors in the probe limit. We note in the discussion section, however, that the structure ``planar AdS$_4$ horizon + hydrodynamic mode = strong IR fluctuations'' should be quite general. In particular, there may be cases in which the metric itself has strong quantum fluctuations over large spacetime distances at any finite value of the AdS radius in Planck units. This would be an interesting and accessible quantum gravitational effect that deserves further study.

\section{Classical symmetry breaking solutions}

While much of our discussion will be general, it will be useful to work through a specific example. We will consider in detail
the case of `holographic superconductors' with 2+1 field theory dimensions
\cite{Gubser:2008px, Hartnoll:2008vx, Hartnoll:2008kx}. A classical treatment of these systems shows spontaneous symmetry
breaking of a $U(1)$ symmetry at temperatures $T < T_C$, as we now briefly review. We will then go on to show how bulk loop
effects will wash out the symmetry breaking expectation value over large distances.

Holographic superconductors are minimally described by a charged scalar field coupled to Einstein-Maxwell theory with a negative
cosmological constant. For our purposes it will be sufficient to work in the probe limit \cite{Hartnoll:2008vx}, in which the Abelian-Higgs
sector does not backreact on the geometry. In this case the background is given by the planar Schwarzschild-AdS spacetime
\be
ds^2 = - f(r) dt^2 + \frac{dr^2}{f(r)} + \frac{r^2}{L^2} \left(dx^2 + dy^2 \right) \,,
\ee
with
\be
f(r) = \frac{1}{L^2} \left(r^2 - \frac{r_+^3}{r} \right) \,,
\ee
where $L$ is the AdS radius and the horizon radius $r_+$ is related to the temperature by $T = 3 r_+/4 \pi L^2$. The Abelian-Higgs model then propagates on this background according to the action
\be\label{eq:abhiggs}
S = - \frac{1}{e^2} \int d^4x \sqrt{-g} \left(\frac{1}{4} F^2 + \left|\pa \Psi - i A \Psi\right|^2 + V(|\Psi|) \right) \,.
\ee
Here $e^2$ is the electromagnetic coupling.
The static solutions of interest to these equations have a nonzero scalar field $\Psi(r)$ (taken to be real in the background solution) and electrostatic potential $\Phi(r) \equiv A_t(r)$. These satisfy
\bea\label{eq:background}
\Phi'' + \frac{2}{r} \Phi'  & = & \frac{2}{f} \Psi^2 \Phi \,, \label{eq:e1} \\ 
\Psi'' + \left(\frac{2}{r} + \frac{f'}{f}\right) \Psi' & = & - \frac{1}{f^2} \Phi^2 \Psi + \frac{1}{2 f} V'(\Psi) \,. \label{eq:e2}
\eea
At this early point for concreteness we will specialise to the specific case
\be
V = m^2 |\Psi|^2 \,.
\ee
The form of the potential will not be important in what follows.

One then looks for solutions to the equations (\ref{eq:e1}) and (\ref{eq:e2}) in which the time component of the Maxwell field asymptotes to
\be
\Phi = \mu + \rho \, \frac{L^2}{r} + \cdots \,.
\ee
Here $\mu$ is the chemical potential which is fixed, while $\rho$ is the charge density that is determined from the equations of motion and regularity at the horizon. It was found in \cite{Hartnoll:2008vx} that there exists a critical temperature $T_C$ such that for $T>T_C$ the only regular solution has vanishing scalar field $\Psi$ while for $T < T_C$ the thermodynamically preferred solution has a nonzero scalar field with asymptotic behaviour
\be\label{eq:vev}
\Psi = \frac{\langle \ocal \rangle}{\sqrt{2}} \frac{1}{r^{3/2 + \sqrt{9/4 + L^2 m^2}}} + \cdots \,.
\ee
This is the `normalisable' behaviour of a massive scalar field in an asymptotically AdS$_4$ spacetime and therefore corresponds to a symmetry breaking expectation value $\langle \ocal \rangle$ of the dual charged operator $\ocal$ in the field theory. By numerically solving the background equations (\ref{eq:e1}) and (\ref{eq:e2}) one can obtain e.g. the condensate $\langle \ocal \rangle(T)$ as a function of temperature. It was checked that the phase transition at $T=T_C$ is second order with mean field exponents. In this paper we will explore the fate of the classical low temperature symmetry broken phase under bulk quantum corrections.

\section{Green's functions for multiple fields}

In the remainder of this paper we will proceed to compute a one loop correction to the classical bulk expectation value $\Psi(r)$, whose asymptotic value determines the dual field theory expectation value $\langle \ocal \rangle$ through (\ref{eq:vev}). A key ingredient that is needed in any one loop computation is of course the propagators of the fluctuating fields. To this end we now obtain the retarded Green's functions for fluctuations about the background. In doing this we develop some machinery for computing curved background propagators for fields that mix.

\subsection{Fluctuation equations in Feynman gauge}

There are in general six real fields to fluctuate: $\{\d A_t, \d A_x, \d A_y, \d A_r, \d \Psi_r, \d \Psi_i\}$, where the last two are the real and imaginary parts of $\delta \Psi$. The fluctuations are both frequency and momentum dependent. Without loss of generality we can take the momentum to be in the $x$ direction so that all fields have the form
\be
\Phi_I(r,x,t) = \Phi_I(r) e^{- i \w t + i k x} \,.
\ee
Here, as we will often do below, we use $\Phi_I$ to denote a generic field fluctuation. It is easily seen that the transverse gauge field modes, $\d A_y$, decouple from the other modes and so can be set to zero for our purposes.

Our theory has the abelian gauge invariance
\be\label{eq:gaugeinvariance}
A \to A + d \lambda\, , \qquad \Psi \to e^{i \lambda}\Psi \, .
\ee
In order to obtain invertible bulk propagators we need to fix the gauge. It will be convenient to use the covariant `$\xi$ gauge', which amounts to adding to the action (\ref{eq:abhiggs}) the term
\be
S_\text{gauge} = - \frac{1}{2 e^2} \frac{1}{\xi} \int d^4x \sqrt{-g} (\nabla_a A^a)^2 \,.
\ee
With this gauge choice, the ghosts decouple from physical fields and will not play a role in our computations.
We will immediately specialise to the Feynman gauge $\xi = 1$ as this simplifies the equations of motion. 

From the gauge fixed action we obtain five second order equations of motion for five variables (we keep $\delta A_y = 0$):
\bea
& & \frac{1}{r^2} (r^2 f \d \Psi_r')' + \left(\frac{\w^2}{f} - \frac{L^2 k^2}{r^2} - m^2 + \frac{1}{f} \Phi^2 \right) \d \Psi_r
+ \frac{2 i \w}{f} \Phi \d \Psi_i + \frac{2}{f} \Phi \Psi \d A_t  = 0 \,, \nonumber \\
& & \frac{1}{r^2} (r^2 f \d \Psi_i')' + \left(\frac{\w^2}{f} - \frac{L^2 k^2}{r^2} - m^2 + \frac{1}{f} \Phi^2 \right) \d \Psi_i
- \frac{2 i \w}{f} \Phi \d \Psi_r  - \frac{1}{r^2 \Psi} \left(r^2 f \Psi^2 \d A_r \right)' \nonumber \\
& & \hspace{1.2cm} - i \Psi \left(\frac{\w}{f} \d A_t + \frac{L^2 k}{r^2} \d A_x \right)  =  0 \,, \nonumber \\
& & - \frac{f}{r^2} (r^2 \d A_t')' - \left(\frac{\w^2}{f} - \frac{L^2k^2}{r^2} - 2 \Psi^2 \right) \d A_t
+ 2 \Psi \left(2 \Phi \d \Psi_r + i \w \d \Psi_i \right)  + i \w f' \d A_r=  0 \,, \nonumber \\
& & (f \d A_x')' + \left( \frac{\w^2}{f} - \frac{L^2 k^2}{r^2} - 2 \Psi^2 \right) \d A_x + 2 i k \left(\Psi \delta \Psi_i + \frac{f}{r} \d A_r \right)  =  0 \,, \nonumber \\
& & \frac{1}{r^2 f} (r^2 f^2 \d A_r')'  + \left( \frac{\w^2}{f} - \frac{L^2 k^2}{r^2} - 2 \Psi^2 + \left(\frac{2f}{r}\right)'+f''  \right) \d A_r  + 2 \Psi^2 \left(\frac{\d \Psi_i}{\Psi} \right)'  \nonumber \\
& & \hspace{1.2cm} - \frac{i \w f'}{f^2} \d A_t - \frac{2 i k L^2}{r^3} \d A_x =  0 \,. \label{eq:eqns}
\eea
These equations were studied in \cite{Amado:2009ts}, in the gauge $\d A_r = 0$ rather than the Feynman gauge, and
the hydrodynamic second sound mode was extracted numerically. We shall rederive some of their results analytically. While the gauge $\d A_r = 0$ is often more convenient for holographic computations, it also introduces a constraint which complicates loop calculations. Mainly for this reason we will stick with the Feynman gauge.

These equations come together with boundary conditions at the horizon and at the boundary. Physical modes should be normalisable at
the boundary and ingoing at the horizon. It is instructive to characterise the asymptotic behaviours precisely. Near the horizon, the general solution takes the form
\bea
& \d A_t  \sim  a_t (r - r_+)^{\nu + 1} \,, \qquad \d A_x \sim a_x (r - r_+)^{\nu} \,, \qquad \d A_r \sim a_r (r - r_+)^{\nu} \,,  & \nonumber \\
& \d \Psi_r  \sim \psi_r (r - r_+)^{\nu} \,, \qquad \d \Psi_i \sim \psi_i (r - r_+)^{\nu} \,. &
\eea
A basis for the five ingoing modes has the following exponents and coefficients
\bea
 \displaystyle \nu = - \frac{i \w}{4 \pi T}  :\, \{a_t, a_x, a_r, \psi_r, \psi_i \} & \propto &   \{- 4 \pi T,0,1,0,0 \} \,,  \{- 2 \Psi(r_+),0,0,0,1 \} \,, \nonumber \\
 & & \{0,1,0,0,0 \} \,, \{0,0,0,1,0 \} \,. \nonumber \\
 \displaystyle \nu = - 1 - \frac{i \w}{4 \pi T}  :\, \{a_t, a_x, a_r, \psi_r, \psi_i \} & \propto & \{ 4 \pi T,0,1,0,0 \} \,. \label{eq:ingoing}
\eea
The five outgoing modes at the horizon have the same form but with $\w \to - \w$ in the above expression for $\nu$ and $a_t \to - a_t$. We will also use the outgoing modes in constructing the Green's function. Recall that $4 \pi T = 3 r_+/L^2$. Note that the $\d A_r = 0$ gauge leads to behaviours at the horizon that cannot be classified as ingoing or outgoing. This is related to the presence of a constraint in this gauge.

Near the boundary, $r \to \infty$, the general solution takes the form
\be
 \d A_t  \sim  a_t r^{\mu} \,, \qquad \d A_x \sim a_x r^{\mu} \,, \qquad \d A_r \sim a_r r^{\mu-2} \,,  \qquad \d \Psi_r  \sim \psi_r r^{\mu} \,, \qquad \d \Psi_i \sim \psi_i r^{\mu} \,.
\ee
A basis for the five normalisable modes now has
\bea
 \displaystyle \mu = - 1 :\, \{a_t, a_x, a_r, \psi_r, \psi_i \} & \propto &   \{e L,0,0,0,0 \} \,,  \{0,e L,0,0,0 \} \,. \nonumber \\
 \displaystyle \mu = - 2  :\, \{a_t, a_x, a_r, \psi_r, \psi_i \} & \propto & \{0,0,e L^3,0,0 \} \,. \label{eq:normalisable} \\
 \displaystyle \mu = - 3/2 - \sqrt{9/4 + L^2 m^2}  :\, \{a_t, a_x, a_r, \psi_r, \psi_i \} & \propto & \{0,0,0, e L^2/\sqrt{2},0 \} \,,  \{0,0,0,0,e L^2/\sqrt{2} \} \,. \nonumber
\eea
The orthonormality of these modes with respect to the appropriate inner product will provide a simplification shortly, we have chosen the above normalisation for future convenience.
A basis for the five non-normalisable modes takes the same form, but with the exponents changed to
\be
\hat \mu = 0\,, \qquad \hat \mu = 1 \,, \qquad \hat \mu = - 3/2 + \sqrt{9/4 + L^2 m^2} \,.
 \label{eq:nonnormalisable}
\ee
The hats are for future use, to distinguish the normalisable from the non-normalisable falloffs.
For simplicity we will always work with the quantisation of the scalar field in which the normalisable mode is that which falls off faster near the boundary.

Having identified the physical modes at the horizon and near the boundary, we can proceed to write down an expression for the retarded Green's function.

\subsection{The retarded Green's function}

The retarded Green's function satisfies
\be\label{eq:Gdef}
K_{IJ} G^R_{JK}(r,r') = \d_{IK} \d(r,r') \,,
\ee
together with ingoing boundary conditions at the horizon and normalisable boundary conditions at infinity.\footnote{We will be defining the retarded Green's function by analytic continuation of (the Fourier transform of) the Euclidean Green's function $G_{IJ}(x,y) = - T_E \langle \Phi_I(x) \Phi_J(y) \rangle$ from the upper half complex frequency plane. In particular this means that there is no $i$ in equation (\ref{eq:Gdef}).} In the above expression $\{I, J, K\}$ run over the five fields and $K_{IJ}$ is the second order differential operator of the equations (\ref{eq:eqns}), written in self-adjoint form, i.e.
\be\label{eq:selfadjoint}
K_{IJ} = \frac{d}{dr} \left(F_{IJ} \frac{d}{dr} \right) + M_{IJ} \,,
\ee
with $F, M$ Hermitian. To put our equations (\ref{eq:eqns}) in this form we need to redefine one of the fields to
\be\label{eq:shift}
\d \widetilde \Psi_i  = \d \Psi_i - \int_{r_+}^r \Psi dr \, \d A_r \,.
\ee
This redefinition will not play a big role in the following, as $\d A_r$ will vanish on the hydrodynamic mode to the order we work.
The explicit values of $F_{IJ}$ are given in the appendix.

In the appendix we derive the following expression for the Green's function of a self-adjoint system of the form (\ref{eq:selfadjoint})
\be\label{eq:mixedG}
G^R_{IJ}(r,r') =
\left\{
\begin{array}{cc}
\Phi^{\text{(in)}, a}_{I}(r) \left(\widetilde W^{-1}\right)_{ab} \overline \Phi^{\text{(bdy)},b}_J(r') & \text{for}\quad r < r' \\ 
\Phi^{\text{(bdy)}, a}_{I}(r) \left( W^{-1}\right)_{ab} \overline \Phi^{\text{(out)},b}_J(r') & \text{for}\quad r > r' 
\end{array} \right. \,,
\ee
where the two `Wronskian' matrices are
\bea\label{eq:wrons}
W^{ab} & = & F_{IJ} \left( \overline \Phi^{\text{(out)},a}_I \frac{d}{dr} \Phi^{\text{(bdy)},b}_J - \frac{d}{dr} \overline \Phi^{\text{(out)},a}_I \Phi^{\text{(bdy)},b}_J   \right) \,, \\
\widetilde W^{ab} & = & F_{IJ} \left(\frac{d}{dr} \overline \Phi^{\text{(bdy)}, a}_I \Phi^{\text{(in)},b}_J  - \overline \Phi^{\text{(bdy)},a}_I \frac{d}{dr} \Phi^{\text{(in)},b}_J \right) \,.
\eea
The overline denotes complex conjugation.
In the above expressions the indices $a,b$ run over the ingoing or outgoing solutions at the horizon $\Phi^{\text{(in)},a}$, $\Phi^{\text{(out)},a}$  and the normalisable solutions $\Phi^{\text{(bdy)},a}$ at the boundary. Note that the complex conjugate of an outgoing solution is ingoing (the complex conjugate satisfies the equations when the operator $K$ acts from the right). In our particular case, there are five ingoing, five outgoing and five normalisable solutions as given in equations (\ref{eq:ingoing}) and (\ref{eq:normalisable}), so the Wronskians are five by five square matrices. They are invertible except at special frequencies, as we shall see shortly. The Wronskians are constants independent of $r$, as we show in the appendix.

The expression (\ref{eq:mixedG}) is a generalisation of the well known form of the Green's function for boundary value problems involving a single function. While the result is unsurprising, the derivation is not completely trivial and we give details in the appendix.

Given that the Wronskians are constant, we can evaluate them in the near boundary region. To do this we need an expression for the ingoing and outgoing wavefunctions in terms of the near boundary wavefunctions.
Because the equations are linear, in the near boundary region we have 
\bea
\Phi_I^{\text{(out)},a} & = & U^{ab} \Phi_I^{\text{(bdy)},b} + V^{ab} \Phi_I^{\text{(n.n.)},b} + \cdots \,, \\
\Phi_I^{\text{(in)},a} & = & \widetilde U^{ab} \Phi_I^{\text{(bdy)},b} + \widetilde V^{ab} \Phi_I^{\text{(n.n.)},b} + \cdots \,,  \label{eq:nearb}
\eea
for constant square matrices $U,V, \widetilde U, \widetilde V$. In this expression $\Phi^{\text{(n.n.)},b}$ are the non-normalisable solutions of (\ref{eq:nonnormalisable}). 
Recall that near the boundary $\Phi_I^{\text{(bdy)},b} \sim \phi_I^b r^{\mu_b}$ and $\Phi_I^{\text{(n.n.)},b} \sim \hat \phi_I^b r^{\hat \mu_b}$ with coefficients and exponents as described in (\ref{eq:nonnormalisable}) and (\ref{eq:normalisable}). Plugging this expression into (\ref{eq:wrons})
gives the leading order behaviour
\be\label{eq:endW}
W^{ab} =  \mp \overline V^{ab} (\hat \mu_b - \mu_b) \,, \qquad \widetilde W^{ab} = \mp \widetilde V^{ba} (\hat \mu_a - \mu_a) \,.
\ee
The lower sign is for the first mode in (\ref{eq:normalisable}), the rest come with the upper sign. In deriving (\ref{eq:endW})
we used the fact that we chose our bases such that $F_{IJ} \phi_I^a \hat \phi_J^c \propto \delta^{ac}$, see the explicit matrix elements $F_{IJ}$ as given in the appendix, as well as the form of the basis in (\ref{eq:normalisable}). As anticipated, the $r$ dependence has cancelled and the Wronskians are constant.

We can see from (\ref{eq:nearb}) that whenever there exists a mode that is ingoing at the horizon and normalisable at the boundary, then $\widetilde V$ will be non invertible. It follows from (\ref{eq:endW}) that the Wronskian is also not invertible. This implies that the Green's function (\ref{eq:mixedG}) will have a pole at e.g. the (in general complex) frequency at which the `on shell' mode appears. This is of course the usual statement that Green's functions have poles at the quasinormal modes of the black hole, see e.g. \cite{Ching:1995tj, Son:2002sd}.

In the following section we proceed to identify the second sound mode and to characterise the corresponding pole in the Green's function.

\section{Second sound mode and pole in Green's function}

\subsection{Low frequency expansion in the far region and dispersion relation}

In this section we exhibit a gapless solution to the fluctuation equations of motion (\ref{eq:eqns}). That is, we are looking for a solution that exists at small frequencies $\w$ and momenta $k$. We expect such a mode to be present as it is the bulk counterpart of the Goldstone mode in the boundary field theory. At $\w = k = 0$ we have the exact solution
\be\label{eq:pure}
\d \Psi_i = \Psi \,,
\ee
with all other fields zero. This zero mode is a large gauge transformation of the background. The gauge transformation parameter, $\lambda$ in (\ref{eq:gaugeinvariance}), is constant and does not vanish at the boundary; it therefore induces a global symmetry transformation in the dual field theory. The mode (\ref{eq:pure}) is however normalisable. It is the normalisable background profile $\Psi$ in the asymptotically AdS spacetime that allows the gapless sound mode to exist, unlike in flat space where the symmetry broken Higgs phase is gapped.
At finite but small $\{\w, k\}$ it sources other physical modes. We can  expand the equations in $\{\w,k\}$ by writing:
\be\label{eq:wkexpand}
\Phi_I = \Phi_I^{(0)} + \w \Phi_I^{(1),\w} + k \Phi_I^{(1),k} + \w^2 \Phi_I^{(2),\w^2}  + \cdots  \,.
\ee
This is not a straightforward expansion to perform, because the near horizon ($r \to r_+$) and low frequency ($\w \to 0$) limits do not commute.
An expansion of the form (\ref{eq:wkexpand}) is only valid in the `far' region $\frac{r}{r_+} - 1 \gg \frac{\w}{T}$, away from the black hole horizon.

Within the far region, we now solve for the various $\Phi_I^{(n)}$ that are sourced perturbatively by  the zeroth order solution (\ref{eq:pure}), requiring normalisability near the boundary. One finds the following  correction at linear order in $\w$ and $k$
\be\label{eq:firstorders}
\d A_t = - i \w \left(1 - \d A_t^\text{hom.} \right) \,, \quad \d A_x = i k \left(1 - \d A_x^\text{hom.} \right) \,, \quad \d \Psi_r = i \w \d \Psi_r^\text{hom.} \,.
\ee
The three `homogeneous' terms are solutions to the $\w=k=0$ equations. These are chosen so that $\d A_t, \d A_x$ and $\d \Psi_r$ are normalisable at infinity. The $\d A_t$ and $\d \Psi_r$ equations are coupled: After setting the non-normalisable mode of $ \d \Psi_r^\text{hom.}$ to zero
and fixing the overall normalisation of $\d A_t$, we are still left with two constants of integration which can be used to ensure matching onto the near horizon region and the associated ingoing or outgoing boundary conditions. Similarly only the overall scale of $\d A_x^\text{hom.}$ must be fixed, leaving a constant of integration for matching. We will not need to perform these matchings here.

The next step is to look at the equations for the $\Phi_I^{(2)}$ fields. For our purposes of extracting the second sound mode we do not need to solve these equations fully. Rather, we will find an obstruction to the existence of the on shell normalisable mode except on a particular dispersion relation (the logic used here is similar to that in \cite{Iqbal:2010eh}). The $\Phi_I^{(2)}$ fields should be normalisable at infinity. Using this information together with the form of the first order solutions in (\ref{eq:firstorders}), one can expand the equations for the perturbations (\ref{eq:eqns}) near infinity. At leading order one finds that satisfying the $\d A_r$ equation requires
\be\label{eq:cs}
\w^2 = c_s^2 k^2 \,, \qquad \text{with} \qquad c_s^2 = \lim_{r \to \infty} \frac{\pa_r \d A_x^\text{hom.}}{\pa_r \d A_t^\text{hom.}} \,.
\ee
This is the second sound mode found numerically in \cite{Amado:2009ts}. That paper also found the expected imaginary part of the dispersion relation which is higher order in $k$. This can be extracted in principle from the equations but will not be necessary here. We can also check that (\ref{eq:cs}) agrees with the (probe limit of the) thermodynamic formula for the second sound derived from relativistic hydrodynamics in \cite{Herzog:2008he}. Recall that the homogeneous solutions are solutions to the $\w=k=0$ equations whose non-normalisable mode tends to one at infinity. The coefficient of the normalisable mode, $\lim_{r \to \infty} \frac{r^2}{L^2} \pa_r \d A_{x/t}^\text{hom.}$, is then precisly the thermodynamic susceptibility for the field theory operator dual to $A_{x/t}$ \cite{Hartnoll:2009sz}. Thus, in the notation of  \cite{Herzog:2008he}
\be\label{eq:thermo}
c_s^2 = - \frac{\left. \pa^2 P/\pa \xi^2 \right|_{T,\mu}}{\left. \pa^2 P/ \pa \mu^2 \right|_{T,\xi}} \,.
\ee
Here $\xi$ is a homogeneous background $A_x$ in the dual field theory. The minus sign is due to the different sign of the bulk kinetic terms for $A_t$ and $A_x.$

In order to find the imaginary part of the dispersion relation we would need to solve the equations in the near horizon region, impose ingoing boundary conditions, and match in the overlap region $\frac{\w}{T} \ll \frac{r}{r_+} - 1 \ll 1$. In our analysis above we noted that there were sufficiently many constants to match onto an ingoing solution at the horizon. However, we could just as well match onto an outgoing solution. The imaginary part of the sound mode pole should have different sign for the retarded and advanced Green's functions (i.e. for ingoing and outgoing boundary conditions). The real, non dissipative, part of the low frequency dispersion relation should be the same for both Green's functions, consistent with what we have found. Via the matching, the ingoing or outoing boundary conditions may source other modes in the far region, in addition to those we have considered, which must vanish as $\w \to 0$. However, because of the linearity of the perturbation equations, these extra terms will not interfere with the terms we have found above.

\subsection{Green's function in the vicinity of the pole}

We found in the previous subsection that with either ingoing or outgoing boundary conditions there is a normalisable on shell mode with low frequency dispersion relation $\w^2 = c_s^2 k^2$. Strictly, we showed that this dispersion is necessary to have a normalisable mode at second order in frequency and momentum. This is sufficient for the following. We noted that the Wronskian matrices will not be invertible at these frequencies. In order to exhibit the pole in the inverse Wronskians we need to find solutions away from the zero mode.

At lowest order in frequency $\w$ we have the zero mode (\ref{eq:pure}). This is the leading order ingoing or outgoing mode and it is also normalisable near the boundary. To obtain the Wronskian matrices via (\ref{eq:endW}) we need to know the corresponding non-normalisable mode generated at higher order in the frequency at the boundary. We noted above that if one substitutes the zeroth (\ref{eq:pure}) and first (\ref{eq:firstorders}) order solutions into the equations of motion (\ref{eq:eqns}) and assumes that the higher order solutions are normalisable, then the equations can only be solved near the boundary when $\w^2 = c_s^2 k^2$. If we take $\w^2 \neq c_s^2 k^2$, then there is a constant term in the $\d A_r$ equation that needs to be cancelled by the addition of a non-normalisable falloff for one of the fields at second order. One finds that the non-normalisable mode that is sourced is $\delta \Psi_i$. Writing as above
$\delta \Psi_i =  \hat \psi_i r^{\hat \mu}$ we solve for the coefficient to find
\be\label{eq:psihat}
 \hat \psi_i  =
- \frac{\sqrt{2}}{\langle \ocal \rangle} \frac{\pa^2 P}{\pa \mu^2} \left(\w^2 - c_s^2 k^2 \right) \,,
\ee
where we used the expressions in (\ref{eq:vev}) and just above (\ref{eq:thermo}) to give the coefficients of zero frequency normalisable modes in terms of field theory quantities. The overall scale ambiguity of the zero mode in (\ref{eq:pure}) will drop out of the Green's function (\ref{eq:mixedG}).

Now recalling from (\ref{eq:nearb}) that e.g. $\Phi^\text{(out)} \sim U \cdot \Phi^{\text{(bdy)}} + V \cdot \Phi^{\text{(n.n.)}}$, we see that the matrices $V$ and $\widetilde V$, and hence the Wronskians (\ref{eq:endW}), have an eigenvalue that is given by the right hand side of (\ref{eq:psihat}). This is because we have found a mode in which both $\Phi_I^{\text{(in/out)}}$ and $\Phi_I^{\text{(n.n.)}}$ in (\ref{eq:nearb}) are proportional near the boundary (to $\delta \Psi_i$), \footnote{We noted previously that matching to the near horizon region will source additional terms in the far region. One might worry that these lead to additional non-normalisable modes. Counting the integration constants, the differential equations allow us to set all the non-normalisable terms to zero except for that of e.g. $\delta \Psi_i$, and also impose ingoing boundary conditions. At $\w=0$ we have an exact solution everywhere (\ref{eq:pure}). At finite $\w$, matching to the near region can only modify the far solution at order $\w$ or higher. In particular, $\delta \Psi_i$ is only modified at order $\w$ or higher. However, we have seen that the order $\w^0$ term in $\delta \Psi_i$ in the far region, together with normalisability, completely determined $\delta A_x$ and $\delta A_t$ at order $\w, k$. These in turn determined the non-normalisable term of $\delta \Psi_i$ at order $\w^2, k^2$. Therefore matching to the near region cannot change the result (\ref{eq:psihat}). At higher order in frequencies we do expect new non-normalisable terms corresponding to the dissipative part of the pole.
} to leading order in the low frequency limit and close to $\w^2 = c_s^2 k^2$. This eigenvalue tends to zero as $\w^2 \to c_s^2 k^2$ and therefore gives the dominant contribution to the inverse matrix. Plugging into (\ref{eq:endW}) and then (\ref{eq:mixedG}) and keeping track of the normalisations in (\ref{eq:ingoing}) and (\ref{eq:normalisable}) gives that the singular term in the Green's function as $\w^2 \to c_s^2 k^2$ is
\be\label{eq:Ganswer}
G^R_{\d\Psi_i \d\Psi_i}(r,r') = e^2 \Theta \frac{\Psi(r) \Psi(r')}{\w^2 - c_s^2 k^2} + \cdots \,,
\ee
where the constant
\be\label{eq:bigtheta}
\Theta = \left(\frac{\pa^2 P}{\pa \mu^2}\right)^{-1} \frac{1}{2 (2 \Delta_\ocal - 3)} \,.
\ee
Here $\Delta_\ocal = 3/2+ \sqrt{9/4 + L^2 m^2}$ is the scaling dimension of the operator $\ocal$ dual to the bulk field $\Psi$.
The expression (\ref{eq:Ganswer}) is valid for all $r,r'$ in the far region and does not have a discontinuity at $r=r'$.
The numerator in (\ref{eq:Ganswer}) will have corrections that go as positive powers of $\w$ or $k$. These will not contribute to the long wavelength divergence that we are aiming for, as will become clear shortly. Similarly, the other components of the Green's function, which we could obtain by turning on different non-normalisable modes, do not have a singularity at $\w^2 = c_s^2 k^2$ and will also not give IR singular fluctuations.

The result (\ref{eq:Ganswer}) is pretty much the expression one might have guessed from the start. Our path has been a little tortuous firstly because we needed to fix the gauge symmetry and secondly because we derived an expression for the Green's function for a general set of coupled fields. We will outline a shorter, if more formal, route to the answer (\ref{eq:Ganswer}) in the discussion section.

\section{Washing out the order parameter}

We can now use the Green's function (\ref{eq:Ganswer}) to compute the one loop correction to the field theory expectation value $\langle \ocal \rangle$. In fact we can do more than this, we will exhibit a one loop IR divergent contribution to the phase of the background field profile $\Psi(r)$, at all values of the radial coordinate $r$.

\subsection{General formalism for one loop corrections}

Consider the following general Euclidean action in a black hole background with a 3 point interaction
\be
S = \int d^4x \sqrt{g(x)} \left( \half \Phi_I K_{IJ} \Phi_J - \Phi_I \Phi_J \Phi_K \lambda^{IJK} \right) \,.
\ee
Here $\lambda^{IJK}$ is symmetric in all indices and can depend on the spacetime coordinates. One can easily generalise the following to allow for a derivative interaction, but this will not be necessary for us. Expanding the path integral in powers of the interaction, the expectation value acquires the one loop tadpole contribution
\be\label{eq:expectation}
\langle\Phi_A(x) \rangle = \Phi_A^\text{cl.} + 3 \int d^4y \sqrt{g(y)} \lambda^{IJK} G_{AI}(x,y) G_{JK}(y,y) + \cdots \,.
\ee
Here $G_{AI}(x,y)$ and so on are the Euclidean Green's function on the black hole background. The Green's function will also have a disconnected part, but these do not contribute to the IR divergence. We will discuss the regime of validity of the expansion in the interaction shortly. Higher point interactions contribute to the tadpole at higher loops and will be subleading in the perturbative regime.

Fourier transform the Green's functions in the field theory directions by writing
\be\label{eq:fourier}
G_{IJ}(x,y) = T \sum_n \int \frac{d^2k}{(2\pi)^2} G_{IJ}(r,r',i\w_n, \vec k) e^{- i \w_n (\t - \t') + i \vec k \cdot (\vec x - \vec x')} \,.
\ee
Here $\tau$ is imaginary time and the Matsubara thermal frequencies for bosons are
\be
\w_n = 2 \pi n T \,, \qquad n \in \Z \,.
\ee
Substituting into (\ref{eq:expectation}) and looking for a homogeneous expectation value in field theory (i.e. with no $\t$ or $\vec x$ dependence) we obtain that the one loop correction is
\be
\langle\Phi_A(r) \rangle =  T \sum_n \int \frac{d^2k}{(2\pi)^2} \int dr' \sqrt{g(r')} \, 3 \lambda^{IJK}
 G_{AI}(r,r',0,0) G_{JK}(r',r', i \w_n,\vec k) \,.
\ee

The next step is to re-express the Matsubara sum in terms of an integral over real frequencies. This is achieved by introducing the contour integral
\be\label{eq:continue}
T \sum_n F(i \w_n) = \frac{-i}{4 \pi} \int_C dz \coth \frac{z}{2T} F(z) \,.
\ee
The contour $C$ has two parts, both running anticlockwise. The first part encircles the poles of $\coth$ along the upper half imaginary axis (including zero) while the second encircles the poles in the lower half plane. The analytic continuation of the Euclidean Green's function depends on whether one continues from the upper or lower imaginary axis. The upper imaginary axis continues into the retarded Green's function, with singularities in the lower half complex frequency plane, while the lower imaginary axis continues into the advanced Green's function, with singularities in the upper half plane. This is easy to see from the boundary conditions at the black hole horizon \cite{Denef:2009yy, Denef:2009kn}. Deforming both parts of the contour $C$ to the real axis gives
\be\label{eq:oneloopfinal}
\langle\Phi_A(r) \rangle  = 3 \int_{-\infty}^{\infty} \frac{d \Omega}{2 \pi} \int \frac{d^2k}{(2\pi)^2} \int dr' \sqrt{g(r')} \coth \frac{\Omega}{2 T} \lambda^{IJK}  \,
   G^R_{AI}(r,r',0,0) \text{Im} \, G^R_{JK}(r',r', \Omega,\vec k) \,. 
 \ee
All of the Green's functions appearing here are retarded Green's functions.
In deriving this expression we used the fact that the symmetric part of the Green's function at real frequencies and momenta satisfies
\be
G^R_{(JK)}(r',r',\Omega,k) - G^A_{(JK)}(r',r',\Omega,k) = 2 i \text{Im} \, G^R_{(JK)}(r',r',\Omega,k) \,.
\ee
This last equality can be derived from the definition of the retarded Green's function in (\ref{eq:mixedG}) using manipulations like those in the appendix. The advanced Green's function $G^A$ is defined via swapping all appearances of $\Phi^{\text{(in)}}$ and $\Phi^{\text{(out)}}$ in $G^R$.
The imaginary part of the retarded Green's function is of course (minus) the spectral density, so it is a natural quantity to appear here.

\subsection{The IR divergence}

We can now use the final expression (\ref{eq:oneloopfinal}) to evaluate the one loop correction to the classical background profile of the scalar field $\Psi$. Using the facts (i) that the singular contribution to the Green's function we are interested in appears in the $\langle \d \Psi_i \d \Psi_i \rangle$ components as in (\ref{eq:Ganswer}) and has imaginary part (using the appropriate $i \epsilon$ prescription for the retarded Green's function)
\be\label{eq:imag}
 \text{Im} \, G^R_{\delta \Psi_i \delta \Psi_i}(r,r, \Omega,\vec k) = - \pi e^2 \Theta \Psi(r)^2 \, \text{sgn}(\Omega)\delta(\Omega^2 - c_s^2 k^2) + \cdots \,,
\ee
(ii) that at $\w=k=0$ the real part of the scalar field $\delta \Psi_r$ only mixes with $\d A_t$, as we see in the equations of motion (\ref{eq:eqns}), and (iii) that expanding the action (\ref{eq:abhiggs}) leads to the following three point coupling involving two $\d \Psi_i$s,
\be
\lambda \equiv \lambda^{\d A_t \d \Psi_i \d \Psi_i} = \frac{2 \Phi(r)}{e^2 f(r)} \,,
\ee
we see that (\ref{eq:oneloopfinal}) implies
\bea
\langle \delta \Psi_r \rangle & = & 3\int_{-\infty}^{\infty} \frac{d \Omega}{2 \pi} \int \frac{d^2k}{(2\pi)^2} \int dr' \sqrt{g(r')} \coth \frac{\Omega}{2 T}  \lambda  \, G^R_{\delta \Psi_r  \delta A_t}(r,r',0,0) \text{Im} \, G^R_{\delta \Psi_r \delta \Psi_r}(r',r', \Omega,\vec k) \nonumber \\
  & = & - \frac{3 \Theta}{2 \pi c_s}  \int dr' G^R_{\delta \Psi_r  \delta A_t}(r,r',0,0) \frac{r'^2 \Psi(r')^2 \Phi(r')}{L^2 f(r')} \int_\Lambda dk \coth \frac{c_s k}{2 T} + \cdots \,, \nonumber \\
 & = & - \Psi(r) \frac{3 e^2 \Theta}{4 \pi c_s} \int_\Lambda dk \coth \frac{c_s k}{2 T} + \cdots \,.   \label{eq:int}
\eea
In the second line we used (\ref{eq:imag}) and also introduced a long wavelength cutoff $\Lambda$ on the momentum integral, which we could think of for instance as a finite volume. As written the integral also has short wavelength divergences, but these can be renormalised away and are not relevant for the low energy physics we are after. To obtain the third line, we used the equation of motion (\ref{eq:Gdef}) for the Green's function $G^R_{\d\Psi_r \d \Psi_r}$, the background equation of motion (\ref{eq:e2}), and integrated by parts twice. A similar argument shows that the electrostatic potential does not receive a divergent quantum correction: $\langle \d A_t \rangle = \text{finite}$.

The IR singular part of (\ref{eq:int}) behaves as
\be\label{eq:thedivergence}
\langle \delta \Psi_r \rangle = - \Psi(r) \frac{3 e^2 \Theta T}{2 \pi c_s^2} \, \log \frac{\mu}{\Lambda} + \cdots \,,
\ee
which has a logarithmic divergence as $\Lambda \to 0$ at all values of the radial coordinate. We replaced an arbitrary renormalisation scale in the logarithm with the chemical potential $\mu$, which is the natural scale in the system at low temperatures.
We can see immediately that any terms with extra positive powers of $\w$ or $k$ appearing in the numerator of the Green's function (\ref{eq:Ganswer}) will lead to extra powers of $k$ in the integral (\ref{eq:int}) and hence do not contribute to the IR singularity. This justifies our neglect of such terms.

The long wavelength divergence (\ref{eq:thedivergence}) is our main result. We should now explain its significance and regime of validity.
Despite superficial appearances, (\ref{eq:thedivergence}) does not imply that the correction to the magnitude of the background scalar field diverges, but rather that the phase is becomingly completely randomised due to wild quantum fluctuations. If the magnitude of the field were diverging, this would be seen in both the real and imaginary parts.\footnote{By working with cylindrical coordinates rather than real and imaginary parts, one sees directly that there is not a divergent tadpole contribution to the magnitude of the field.} We see above that only the real part of the field has divergent fluctuations. The classical expectation value was chosen to be real; therefore as the phase fluctuates around in a circle the positive and negative fluctuations of the scalar field in the imaginary direction cancel out while there is a net fluctuation in the real part. Alternatively: if the magnitude of the field does not diverge, but the real part is becoming ill-defined, it must be because the phase is not defined. Thus while the classical solution for the background magnitude of the scalar field receives only small quantum corrections, the phase does not have a specific value and spontaneous symmetry breaking does not occur.

For perturbation theory to be applicable, the electromagnetic coupling should be small $e \ll 1$. The one loop correction in (\ref{eq:thedivergence}) is down by a factor of $e^2$ compared to the classical value. This remains true so long as the IR momentum cutoff $\Lambda$, which we might think of as defining a spatial volume Vol, satisfies
\be
\text{Vol}^{1/2} \equiv \Lambda^{-1} \ll \Lambda_\text{fluc.}^{-1} \equiv \mu^{-1} e^{\#/e^2}\,.
\ee
Here $\#$ is an order one positive number. Thus, over exponentially large (in $e^{-2} \gg 1$) volumes of space, in the dual field theory directions $\{x,y\}$, perturbation theory is reliable and quantum corrections are small. However, on nonperturbatively large volume scales, quantum fluctuations eventually dominate and restore the broken symmetry. While the one loop correction itself is no longer reliable on these scales, the appearance of the dynamical scale
$\Lambda_\text{fluc.}$ should be interpreted analogously to e.g. the perturbative appearance of the QCD scale $\Lambda_\text{QCD}$: it is the scale at which a logarithmically running effect becomes order one and qualitatively transforms the physics. In our present case it is clear that the physics in question is the randomization of the phase.

In our probe limit, the overall scale of the action has been set by $e$. However, once backreaction onto the metric is incorporated, the magnitude of the on shell action is more typically set by some positive power of `$N$', where `$N$' measures the degrees of freedom `per site' in the dual field theory. In practice, from a bulk point of view, $N$ is some inverse power of the AdS radius in Planck units. The limit in which gravitational perturbation theory is valid becomes $N \gg 1$, and the lengthscale above which quantum fluctuations dominate the geometry becomes
\be
\Lambda_\text{fluc.}^{-1} \equiv \mu^{-1} e^{\# N^{\#}}\,.
\ee

A potential limitation of the regime of validity of our result comes from the fact that we only obtained the Green's function in the far region, $\frac{\w}{T} \ll \frac{r}{r_+} - 1$. Near the horizon, at $\frac{r}{r_+} \lesssim 1 + \frac{\w}{T}$, the expression (\ref{eq:imag}) cannot be used because the small $\w$ expansion of (\ref{eq:wkexpand}) is not valid there. However, in the hydrodynamic limit this non-far region gives a vanishingly small contribution to the integral over $r'$ in (\ref{eq:int}). This fact combined with the above observation that in the far region the integrand is effectively a delta function $\delta(r-r')$ implies that the answer (\ref{eq:thedivergence}) should be true arbitrarily close to the horizon. That is, for any given $r$ the dominant values of $\w$ and $k$ are sufficiently small that our result holds.

Before moving on to discuss the presence of algebraic long range order, note a small but interesting difference between the computation we have just performed and the more usual computation involving a weakly interacting scalar field in 2+1 dimensional flat space with a mexican hat type of potential. There, the interaction that leads to a divergence like (\ref{eq:thedivergence}) comes from the form of the scalar potential. In our case we set the potential to be simply a mass term and so there are no such interactions. Instead, we found that the important interaction involved the bulk Maxwell field. This Maxwell-scalar interaction is not present in the usual field theoretic case, as the symmetry has to be global. This difference highlights one sense in which the bulk computation does not fit into the conventional statement of the CMWH theorem.

\subsection{Algebraic long range order}

Our computations so far have shown how, at the level of one point functions, quantum fluctuations randomise the phase of the scalar field over large distances and restore the classically broken symmetry. Ordered phases are more constructively characterised by the existence of `off-diagonal long range order'. This is the statement that the equal time spatial correlator of the order parameter does not vanish at large separations
\be
\lim_{| \vec x| \to \infty} \langle \ocal(\vec x)^\dagger \ocal(0) \rangle \neq 0 \,.
\ee
In disordered phases, in contrast, the correlator vanishes exponentially at large separations, with the correlation length setting the scale beyond which the correlations are small. In 2+1 dimensional theories to which the CMWH theorem applies, one typically finds that the low temperature phase that is classically ordered has instead `algebraic long range order'.  We now proceed to show that our holographic system exhibits algebraic long range order, both in the bulk and in the dual boundary field theory. Specifically we will find
\be\label{eq:algebraic}
\lim_{| \vec x| \to \infty} \langle \ocal(\vec x)^\dagger \ocal(0) \rangle \sim \frac{1}{|\vec x|^{\# e^2}} \,.
\ee
Thus in the strict classical limit $e \to 0$, genuine long range order is recovered. For any finite $e$ the correlation eventually goes to zero at sufficiently large distances. This situation is very analogous to that discussed in \cite{Witten:1978qu}. At shorter distance scales, which at small $e$ can be much larger than $\mu$, the field is locally ordered. In this regime previous results on holographic superconductors in 2+1 dimensions will carry over. At larger values of $e$ the classical results will be misleading.

The ingredients going into (\ref{eq:algebraic}) are closely related to those leading to our previous result (\ref{eq:thedivergence}) and so we can be brief. We noted in the previous section that the magnitude of the scalar field only receives small quantum corrections, while the phase fluctuates strongly. It is therefore useful to write the field in polar form $\Psi = \Psi(r) e^{i \theta}$, with $\Psi(r)$ the classical background profile and $\theta$ the phase. The equal time correlator of the order parameter can then be written, to leading order in perturbation theory in $e$,
\be\label{eq:thetatheta}
\langle \Psi(r,\vec x)^\dagger \Psi(r',0) \rangle = \Psi(r) \Psi(r') \langle e^{- i (\q(r, \vec x) - \q(r',0)) } \rangle = \Psi(r) \Psi(r') e^{ - \half \langle (\q(r, \vec x) - \q(r',0))^2 \rangle}\,.
\ee
To obtain the second equality one performs a standard manipulation using Wick contractions.

We have already computed the two point correlators of the phase appearing in (\ref{eq:thetatheta}). For purposes of computing the two point function about a background where the scalar is real, fluctuations of the phase are given by fluctuations of the imaginary part of the field considered previously: $\delta \Psi_i = \Psi(r) \, \theta$. Thus from (\ref{eq:Ganswer}) the retarded Green's function for the phase has the singular behaviour
\be\label{eq:thetasing}
G^R_{\theta\theta}(r,r') = \frac{e^2 \Theta}{\w^2 - c_s^2 k^2} + \cdots \,.
\ee
Note that this singular term does not depend on $r,r'$. Using this fact as well as the Fourier transformation of (\ref{eq:fourier}) and the analytic continuation technique of (\ref{eq:continue}), the exponent of the correlation function (\ref{eq:thetatheta}) becomes
\be
\langle \theta(\vec x) \theta(0) \rangle - \langle \theta(0) \theta(0) \rangle = - \int_{-\infty}^{\infty} \frac{d \Omega}{2 \pi} \int \frac{d^2k}{(2\pi)^2} \coth \frac{\Omega}{2 T}    \, \text{Im} \, G^R_{\theta \theta}(\Omega,\vec k) \left(e^{i \vec k \cdot \vec x} - 1 \right) \,. 
\ee
In the limit of large separation $|\vec x| \to \infty$, the dominant contribution to the correlation comes from the IR singular region of (\ref{eq:thetasing}). To leading order at large separation the integrals above can be evaluated to give
\be
\lim_{| \vec x| \to \infty} \left( \langle \theta(\vec x) \theta(0) \rangle - \langle \theta(0) \theta(0) \rangle \right) = - \frac{e^2 \Theta T}{2 \pi c_s^2} \log \mu | \vec x|  + \cdots \,.
\ee
We again inserted $\mu$ as a characteristic scale.
Putting the above together we obtain the algebraic long range order
\be
\lim_{| \vec x| \to \infty} \langle \Psi(r,\vec x)^\dagger \Psi(r',0) \rangle = \Psi(r) \Psi(r') |\vec x|^{-\frac{e^2 \Theta T}{2 \pi c_s^2}} + \cdots \,.
\ee
Taking $r$ and $r'$ to the boundary and scaling out the conformal factor, we obtain the advertised dual field theory result (\ref{eq:algebraic}).

\section{Discussion}

In this paper we have shown how the quantum fluctuations of a hydrodynamic second sound mode in a hairy black hole spacetime wash out the classical symmetry breaking phase. This effect was expected as a consequence of the CMWH theorem in the dual field theory. Technically our computation depended on identifying the second sound mode, isolating the contribution of this mode to the Green's function, and then using standard manipulations from finite temperature quantum field theory. These techniques also apply in black hole backgrounds due to the boundary conditions at the horizon.

Before ending with some comments on future directions, we will outline an alternative, more direct, route towards obtaining the singular part of the Green's function. This may be useful in more general contexts, e.g. with gravitational backreaction, where many modes are coupled. However, the construction is a little formal and should be scrutinised carefully before use. The idea, inspired by analogous results for determinants \cite{Denef:2009yy, Denef:2009kn}, is to express the Green's function as a sum over quasinormal modes. In terms of the eigenvalues to the self-adjoint Euclidean operator
\be
\frac{d}{dr} \left(F_{IJ} \frac{d}{dr} \Phi^{(\a)}_J \right) + M_{IJ}(i \w_n,k) \Phi^{(\a)}_J  =  \lambda_\a(i \w_n,k) \Phi^{(\a)}_I \,,
\ee
the Euclidean Green's function may be written as
\be
G_{IJ}(r,r'; i \w_n,k) = \sum_\a \frac{\overline \Phi^{(\a)}_I(r; i \w_n,k)  \Phi^{(\a)}_J(r'; i \w_n,k) }{\l_\a(i \w_n,k)} \,.
\ee
In this expression the eigenfunctions must be taken to be orthonormal.
We can now view the Green's function as a meromorphic function of complex frequency, analytically continuing from $\w = i \w_n$ on the upper imaginary axis to obtain the retarded Green's function. Up to `local' UV terms (cf. \cite{Denef:2009kn}) the function will be determined by the residues of its poles in the complex $\w$ plane. These poles occur at frequencies $\w_\star(k)$ at which an eigenvalue vanishes $\lambda(\w_\star(k),k)=0$. These are the quasinormal frequencies of the black hole. At these frequencies, the eigenfunction $\Phi^{(\a)}_I(r; \w_\star(k),k)$ becomes the corresponding quasinormal mode $\Phi^\star_I(r)$. The retarded Green's function may therefore be written as
\be
G_{IJ}^R(r,r';\w,k) = \sum_{\w_\star(k)} \left.\frac{ d\lambda}{d\w} \right|^{-1}_{\w=\w_\star} \frac{\overline \Phi^\star_I(r) \Phi^\star_J(r')}{\w - \w_\star(k)} \,.
\ee
If, as in this paper, we are interested in the IR physics due to a hydrodynamic quasinormal mode, then we need retain only this term in the above sum over quasinormal modes and furthermore in the numerator we can put $\w=k=0$. In our system, the quasinormal modes at $\w_\star = \pm c_s k$ together with the eigenvalue $\left. \pa_\w \lambda \right|_{\w=\w_\star} \sim \mp k$ leads to our previously obtained form (\ref{eq:Ganswer}). Matching the numerical prefactor requires more work, in particular grappling with the correct normalisation of the quasinormal modes, and we do not pursue this here.

An intriguing possibility is that there may be a proliferation of IR singular quantum effects in curved spacetime backgrounds in 3+1 dimensions with horizons. These need not be limited to asymptotically AdS spacetimes or to strictly planar horizons. It may be interesting to explore cosmological applications \cite{Tsamis:1996qq, Polyakov:2007mm}. The `membrane paradigm' approach to horizons may naturally provide the necessary light hydrodynamic modes. Similarly to what we have seen in this paper, sending these modes around quantum loops may lead to effects familiar from 2+1 dimensional finite temperature physics in flat space. For instance, the shear viscosity diverges in 2+1 dimensions \cite{divergence}.

Finally, while we have examined the quantum physics of the low temperature phase, we have not looked at the phase transition itself. It seems likely that a bulk renormalisation group analysis close to $T = T_C$ will show a Berezinskii-Kosterlitz-Thouless (BKT) transition \cite{bkt} rather than the classical second order phase transition. It would be nice to exhibit various BKT scalings explicitly.

\section*{Acknowledgements}

This work has benefitted from enjoyable discussions with Frederik Denef, Tom Hartman, Hong Liu, Mark Mezei,  Omid Saremi and Dam Son. D.A. would especially like to thank Georgios Pastras for initial collaboration on related topics. Jennifer Marsh helped to translate parts of Bloch's 1930 paper. Our research is partially supported by DOE grants DE-FG02-91ER40654 and DE-FG02-05ER41360 and (S.A.H.) the FQXi foundation.

\appendix

\section{Green's function for one dimensional boundary value problem with multiple fields}

In this appendix we derive the formula (\ref{eq:mixedG}) for the Green's function as well as the statement that the `Wronskians' (\ref{eq:wrons}) are constant. That is, we will obtain the Green's function for a one dimensional boundary value problem involving multiple fields.

The Green's function satisfies
\be\label{eq:appendG}
\left( \frac{d}{dr} \left(F_{IJ} \frac{d}{dr} \right) + M_{IJ} \right) G^R_{JK}(r,r') = \d_{IK} \d(r,r') \,,
\ee
with $F, M$ Hermitian matrices, together with boundary conditions. To tie in with the main text, we will call the boundary conditions at $r=r_+$, `ingoing/outgoing', and the boundary condition at $r \to \infty$, `normalisable'. Also, we will assume that if there are $N$ fields, i.e. $I = 1 \ldots N$, then there are $N$ ingoing boundary modes and $N$ normalisable modes. We will also need the $N$ outgoing modes whose complex conjugates are ingoing. We can therefore define the $N \times N$ square matrices
\be
\Phi_\text{in/out} \equiv \Phi^{\text{(in/out)}}_{a \, I} \,, \qquad\Phi_\infty \equiv \Phi^{\text{(bdy)}}_{a \, I} \,,  \qquad G^R \equiv G^R_{IJ} \,, \qquad P \equiv P_{ab} \,, \qquad Q \equiv Q_{ab} \,. 
\ee
Here the matrices $\Phi^{\text{(in/out)}}_{a \, I}$ and $\Phi^{\text{(bdy)}}_{a \, I}$ combine the $N$ field components of the $N$ ingoing, outgoing and normalisable solutions, each labelled by $a=1 \ldots N$. We then make the following ansatz for the Green's function
\be
G^R(r,r') =
\left\{
\begin{array}{cc}
\Phi^T_\text{in}(r) P(r') \overline \Phi_\infty(r') & \text{for}\quad r < r' \\ 
\Phi^T_\infty(r) Q(r') \overline \Phi_\text{out}(r')  & \text{for}\quad r > r' 
\end{array} \right. \,,\label{eq:ans}
\ee
We now proceed to solve for $P, Q$ and to show that they are in fact constant and independent of $r'$. As in the text, overline denotes complex conjugation.

There are two matching conditions to impose at $r=r'$. The fields must be continuous and the derivative of the fields must jump according to (\ref{eq:appendG}). Continuity is immediately seen to imply
\be
P = (\Phi_\infty \Phi_\text{in}^{-1})^T Q \overline \Phi_\text{out} \overline \Phi_\infty^{-1} \,.
\ee
The discontinuity of the derivative at $r=r'$ then implies, using the previous continuity equation to eliminate $P$,
\be\label{eq:jump}
\overline \Phi_\text{out} F \left(\frac{d \Phi_\infty^T}{dr} - \frac{d \Phi_\text{in}^T}{dr} \Phi_\text{in}^{-1 \, T} \Phi_\infty^T \right) Q = \text{Id} \,.
\ee
Let us introduce the two matrices
\bea
S & = & \overline \Phi_\text{out} F \frac{d \Phi_\text{in}^T}{dr} - \frac{d \overline \Phi_\text{out}}{dr} F \Phi_\text{in}^T \,, \\
W & =  & \overline \Phi_\text{out} F \frac{d \Phi_\infty^T}{dr} - \frac{d \overline \Phi_\text{out}}{dr} F  \Phi_\infty^T    \,,
\eea
and thereby rewrite (\ref{eq:jump}) as
\be
\left(-S \Phi_\text{in}^{-1 \, T} \Phi_\infty^T + W \right) Q = \text{Id} \,.
\ee
The advantage of this expression is that we will now show that $W$ is constant and that $S$ vanishes. Given that $W$ is the first of the Wronskians as defined in (\ref{eq:wrons}) in the main text, this gives the result we are seeking, $Q = W^{-1}$. A few steps of algebra, using the vanishing of
\be
\widetilde S = \overline \Phi_\infty F \frac{d \Phi_\infty^T}{dr} - \frac{d \overline \Phi_\infty}{dr} F \Phi_\infty^T \,, 
\ee
 then show that $P = \widetilde W^{-1}$. Here
\be
\widetilde W = \frac{d\overline \Phi_\infty}{dr} F \Phi^T_\text{in} - \overline \Phi_\infty F \frac{d \Phi^T_\text{in}}{dr} \,,
\ee
completing the derivation of (\ref{eq:mixedG}).

Constancy of $W, \widetilde W$ and $S, \widetilde S$ is immediate from acting with $\frac{d}{dr}$ and using the Hermiticity of $F$ and $M$. Given that $S$ and $\widetilde S$ are constant, to show that they vanish it is sufficient to show that they vanish at any point. The natural points to take are the horizon for $S$ and infinity for $\widetilde S$. For $\widetilde S$, vanishing follows immediately from the orthogonality of the normalisable modes in (\ref{eq:normalisable}). For $S$ we must work a little harder. Firstly, taking into account the shift (\ref{eq:shift}), we have
\bea
& e^2 F_{tt} = - \frac{r^2}{L^2} \,, \qquad e^2 F_{xx} = f \,, \qquad e^2 F_{\Psi_r \Psi_r} = \frac{2 r^2 f}{L^2} \,, \qquad e^2 F_{\widetilde \Psi_i \widetilde \Psi_i} = \frac{2 r^2 f}{L^2} \,,  & \nonumber \\
&  e^2 F_{rr} = \frac{r^2 f^2}{L^2} + \frac{2 r^2 f q^2}{L^2} \left(\int_{r_+}^r \Psi dr \right)^2 \,, \qquad e^2 F_{r \widetilde \Psi_i} = \frac{2 r^2 f q}{L^2} \int_{r_+}^r \Psi dr \,. &
\eea
Using the ingoing modes of (\ref{eq:ingoing}) we can compute $S$ on each pair of modes. It is found that all components of $S$ vanish at $r=r_+$. This vanishing depended on the combination of $\Phi^T_\text{in}$ and $\overline \Phi_\text{out}$ appearing in our ansatz (\ref{eq:ans}). The combination can be deduced from first principles by working in Euclidean signature, where there is a Dirichlet boundary condition at the horizon, and then analytically continuing to obtain the retarded Green's function.

\qed

The vanishing of $S$ and $\widetilde S$ is presumably general for self-adjoint boundary value problems, as it is necessary for the Green's function to have the required symmetry properties under $r \leftrightarrow r'$. For Dirichlet boundary conditions, for instance, supposing the Wronskians are constant and finite, then the $S$ and $\widetilde S$ matrices must be constant and zero because they have a faster falloff near the relevant boundary.


\begin{thebibliography}{99}

\bibitem{sym}
S.~Coleman,
``There are no Goldstone bosons in two dimensions,''
Commun. Math. Phys. {\bf 31}, 259 (1973).

N.~D.~Mermin and H.~Wagner,
``Absence of ferromagnetism or antiferromagnetism in
one-  or two- dimensional isotorpic heisenberg models,''
Phys. Rev. Lett. {\bf 17}, 1133 (1966).

P.~C.~Hohenberg,
``Existence of long-range order in one and two dimensions,''
Phys. Rev. {\bf 158}, 383 (1967).

\bibitem{bloch}
F.~Bloch, ``Zur theorie des ferromagnetismus,''
Z. Phys. {\bf 61}, 206 (1930). 

\bibitem{Maldacena:1997re}
  J.~M.~Maldacena,
  ``The large N limit of superconformal field theories and supergravity,''
  Adv.\ Theor.\ Math.\ Phys.\  {\bf 2}, 231 (1998)
  [Int.\ J.\ Theor.\ Phys.\  {\bf 38}, 1113 (1999)]
  [arXiv:hep-th/9711200].
  
\bibitem{Klebanov:1999tb}
  I.~R.~Klebanov and E.~Witten,
  ``AdS/CFT correspondence and symmetry breaking,''
  Nucl.\ Phys.\  B {\bf 556}, 89 (1999)
  [arXiv:hep-th/9905104].
  
\bibitem{Denef:2009yy}
  F.~Denef, S.~A.~Hartnoll and S.~Sachdev,
  ``Quantum oscillations and black hole ringing,''
  arXiv:0908.1788 [hep-th].
  
\bibitem{CaronHuot:2009iq}
  S.~Caron-Huot and O.~Saremi,
  ``Hydrodynamic Long-Time tails From Anti de Sitter Space,''
  arXiv:0909.4525 [hep-th].
  
\bibitem{Hartnoll:2009kk}
  S.~A.~Hartnoll and D.~M.~Hofman,
  ``Generalized Lifshitz-Kosevich scaling at quantum criticality from the
  holographic correspondence,''
  arXiv:0912.0008 [cond-mat.str-el].
  
\bibitem{Faulkner:2010da}
  T.~Faulkner, N.~Iqbal, H.~Liu, J.~McGreevy and D.~Vegh,
  ``From black holes to strange metals,''
  arXiv:1003.1728 [hep-th].
  
\bibitem{Hartman:2010fk}
  T.~Hartman and S.~A.~Hartnoll,
  ``Cooper pairing near charged black holes,''
  arXiv:1003.1918 [hep-th].

\bibitem{Hartnoll:2009sz}
  S.~A.~Hartnoll,
  ``Lectures on holographic methods for condensed matter physics,''
  Class.\ Quant.\ Grav.\  {\bf 26}, 224002 (2009)
  [arXiv:0903.3246 [hep-th]].

\bibitem{McGreevy:2009xe}
  J.~McGreevy,
  ``Holographic duality with a view toward many-body physics,''
  arXiv:0909.0518 [hep-th].
  
\bibitem{Witten:1978qu}
  E.~Witten,
  ``Chiral Symmetry, The 1/N Expansion, And The SU(N) Thirring Model,''
  Nucl.\ Phys.\  B {\bf 145}, 110 (1978).
  
\bibitem{Marolf:1999uq}
  D.~Marolf and A.~W.~Peet,
  ``Brane baldness vs. superselection sectors,''
  Phys.\ Rev.\  D {\bf 60}, 105007 (1999)
  [arXiv:hep-th/9903213].
  
\bibitem{Peet:1999bc}
  A.~W.~Peet,
  ``Baldness/delocalization in intersecting brane systems,''
  Class.\ Quant.\ Grav.\  {\bf 17}, 1235 (2000)
  [arXiv:hep-th/9910098].
  
\bibitem{Gregory:2009fj}
  R.~Gregory, S.~Kanno and J.~Soda,
  ``Holographic Superconductors with Higher Curvature Corrections,''
  JHEP {\bf 0910}, 010 (2009)
  [arXiv:0907.3203 [hep-th]].
  
\bibitem{Gubser:2008px}
  S.~S.~Gubser,
  ``Breaking an Abelian gauge symmetry near a black hole horizon,''
  Phys.\ Rev.\  D {\bf 78}, 065034 (2008)
  [arXiv:0801.2977 [hep-th]].

\bibitem{Hartnoll:2008vx}
  S.~A.~Hartnoll, C.~P.~Herzog and G.~T.~Horowitz,
  ``Building a Holographic Superconductor,''
  Phys.\ Rev.\ Lett.\  {\bf 101}, 031601 (2008)
  [arXiv:0803.3295 [hep-th]].

\bibitem{Hartnoll:2008kx}
  S.~A.~Hartnoll, C.~P.~Herzog and G.~T.~Horowitz,
  ``Holographic Superconductors,''
  JHEP {\bf 0812}, 015 (2008)
  [arXiv:0810.1563 [hep-th]].
  
\bibitem{Amado:2009ts}
  I.~Amado, M.~Kaminski and K.~Landsteiner,
  ``Hydrodynamics of Holographic Superconductors,''
  JHEP {\bf 0905}, 021 (2009)
  [arXiv:0903.2209 [hep-th]].

\bibitem{Ching:1995tj}
  E.~S.~C.~Ching, P.~T.~Leung, W.~M.~Suen and K.~Young,
  ``Wave propagation in gravitational systems: Late time behavior,''
  Phys.\ Rev.\  D {\bf 52}, 2118 (1995)
  [arXiv:gr-qc/9507035].

\bibitem{Son:2002sd}
  D.~T.~Son and A.~O.~Starinets,
  ``Minkowski-space correlators in AdS/CFT correspondence: Recipe and
  applications,''
  JHEP {\bf 0209}, 042 (2002)
  [arXiv:hep-th/0205051].

\bibitem{Iqbal:2010eh}
  N.~Iqbal, H.~Liu, M.~Mezei and Q.~Si,
  ``Quantum phase transitions in holographic models of magnetism and
  superconductors,''
  arXiv:1003.0010 [hep-th].
  
\bibitem{Herzog:2008he}
  C.~P.~Herzog, P.~K.~Kovtun and D.~T.~Son,
  ``Holographic model of superfluidity,''
  Phys.\ Rev.\  D {\bf 79}, 066002 (2009)
  [arXiv:0809.4870 [hep-th]].

\bibitem{Denef:2009kn}
  F.~Denef, S.~A.~Hartnoll and S.~Sachdev,
  ``Black hole determinants and quasinormal modes,''
  arXiv:0908.2657 [hep-th].

\bibitem{Tsamis:1996qq}
  N.~C.~Tsamis and R.~P.~Woodard,
  ``Quantum Gravity Slows Inflation,''
  Nucl.\ Phys.\  B {\bf 474}, 235 (1996)
  [arXiv:hep-ph/9602315].
  
\bibitem{Polyakov:2007mm}
  A.~M.~Polyakov,
  ``De Sitter Space and Eternity,''
  Nucl.\ Phys.\  B {\bf 797}, 199 (2008)
  [arXiv:0709.2899 [hep-th]].

\bibitem{divergence}
D.~Forster, D.~R.~Nelson and M.~J.~Stephen,
``Large-distance and long-time properties of a randomly stirred fluid,''
Phys. Rev. {\bf A16}, 732 (1977).

\bibitem{bkt}

V.~L.~Berezinskii,
Zh. Eksp. Teor. Fiz. {\bf 59}, 907 (1970) [Sov. Phys. JETP {\bf 32}, 493 (1971)].

J.~M.~Kosterlitz and D.~J.~Thouless, ``Ordering, metastability and
phase transitions in two-dimensional systems,''
J. Phys. {\bf C6}, 1181 (1973).

\end{thebibliography}
\end{document}